\documentstyle[12pt]{ioplppt}

\begin{document}

\jl{3}

\title{Interpretation 
of Photoemission Spectra of $(TaSe_4)_2I$
as Evidence of Charge Density Wave Fluctuations
}[\JPCM {\bf 8} (1996) 10493--10509 ]

\author{Nic Shannon}
\address{
Department of Physics, 
University of Warwick,
Coventry,
England} 
\author{Robert Joynt}
\address{
Department of Physics, 
University of Wisconsin-Madison\\
1150 University Avenue, 
Madison, Wisconsin 53706\\
and \\
Department of Technical Physics\\
Helsinki University of Technology\\
02150-Espoo, Finland}

\begin{abstract}
The competition between different and unusual effects in 
quasi-one-dimensional conductors makes the direct interpretation of 
experimental measurements of these materials both difficult and 
interesting.  We consider evidence 
for the existence of large charge-density-wave fluctuations
in the conducting phase of the Peierls insulator 
(TaSe$_4$)$_2$I, by comparing the predictions of a simple 
Lee, Rice and Anderson theory for such a system with recent 
angle-resolved photoemission spectra.  The agreement obtained
suggests that many of the unusual features of these spectra
may be explained in this way.  This view of the system is contrasted 
with
the behaviour expected of a Luttinger liquid.
\end{abstract}


\maketitle

\section{Introduction}

Restriction of a many body system to low dimension brings 
with it simplifications but also a cost - the need to pay much closer
attention to the effects of both thermodynamic fluctuations and  
interparticle correlations.  One-dimensional models have proved an 
extremely rich
field for the exploration of subtle effects of electron-electron 
interactions. 
They have been found to have quite complex ground states brought about 
by the same reduced phase space which makes some of their properties 
soluble, and 
which causes finite temperature to play so dramatic a role. 

A colourful
variety of quasi-one- and 
two-dimensional materials is now accessible to experiment,
but this has not produced scientific consensus in all areas.
Nine years of 
intensive research on high-temperature superconductors, the best known 
examples of quasi-two-dimensional systems, have failed to 
yield agreement on even
the symmetry of their respective ground states.

The study of quasi-one-dimensional conductors has been less intense, but
they are interesting for similar reasons.  All attempts to model these 
materials
must make reference to the fact that they are not truly 
one-dimensional, but merely very, very anisotropic.  The set of 
materials 
to which $(TaSe_4)_2I$ belongs have a common structure of an 
assembly of weakly coupled conducting chains.
Their behaviour has generally been understood in terms of those 
properties of one 
(or three) dimensional systems which are held to be relevant to them.  
This interplay of dimensionality is not trivial; weak interchain 
coupling will 
act to stabilise states born of instabilities in the underlying 
one-dimensional structure, which would otherwise be destroyed at
finite temperature by large thermodynamic fluctuations.  In the limit of 
strong coupling both the instability and the fluctuations may be
irrelevant.

Many and diverse experimental techniques have 
been brought to bear on quasi-one-dimensional 
conductors in pursuit of insight 
into their properties and the low dimensional physics which these 
betray.  The transition-metal tetrachalcogenide $(TaSe_4)_2I$ alone, 
in the thirteen years since 
its first synthesis \cite{gressier1}, has been probed with 
neutron scattering \cite{fujishita1}, X-rays \cite{fujishita2},
low energy electrons \cite{rocau}, ultraviolet photons \cite{sato}, 
and subjected 
to measurements of optical and electrical 
conductivity \cite{geserich,berner,wang} and magnetic susceptibility. 
\cite{johnston}.  
This material is interesting, since on cooling it displays 
a second order phase transition at the (relatively high) temperature of 
263 K, 
from a highly 
conductive but extremely anisotropic "metallic" phase into a
semi-conducting 
charge density 
wave ground state.   While angle integrated photoemission
\cite{dardel1,dardel2} and inverse 
photoemision 
\cite{purdie} experiments have been performed on such materials for some 
time, 
only very
recently has angle resolved photoemission (ARPES) data become 
available.  This offers the most direct means yet of understanding the 
changes 
which take 
place in the system's electronic properties.

The instability of one-dimensional metals against an insulating ground 
state was described by 
Peierls \cite{peierls} in 1953 and the mean field theory 
\cite{rice} 
can be constructed in close analogy to the BCS theory of
superconductivity.  
It is also well known that 
the fluctuations forced by the restricted phase space of a truly one-dimensional 
system prevent it from undergoing a phase 
transition at any finite temperature.  The real systems, then, must                                                                   
exhibit an interplay of dimensional and thermodynamic effects, which 
manifest themselves in the strong and unusual temperature 
dependence of the "metallic" properties observed above the 
transition temperature.  Early measurements of the magnetic 
susceptibility of 
$(TaSe_4)_2I$ \cite{johnston} 
suggested that fluctuations of the charge density wave (CDW) 
order parameter were present for all temperatures above 
263K, up to the limit of the compound's 
chemical stability (at about 430K).  In order to understand 
the phase transition which takes place in this 
and other similar materials, it is necessary first to 
understand the role of fluctuations in the properties of a 
one-dimensional 
conductor.

It has also been known for some time that the Landau Fermi liquid state 
is 
unstable against interparticle interaction in one dimension; the 
paradigm 
of the the Luttinger liquid has evolved to describe the properties of a 
number 
of abstract one-dimensional 
models (for an overview see \cite{schultz}). Luttinger liquids have 
unusual 
correlation functions displaying separation of spin and charge degrees 
of freedom, 
and possess no stable single particle excitations at the Fermi energy 
$\epsilon_F$.  
This implies that the 
customary discontinuity in zero temperature occupation 
number $n(k)$ at the Fermi wavevector $k_F$ found in all Fermi liquids 
is 
absent, and that the electronic density of states will 
vanish near $\epsilon_F$, with a power law 
behaviour determined by the strength of interparticle interactions.  The 
decoupling of spin and charge degrees of freedom manifests itself in new 
structure in the spectral function of the system.

A topic much discussed in recent years is whether some aspects of this
well-established one-dimensional behavior can survive in higher 
dimensionalities.
For example, it has been proposed that the normal state properties of
high-temperature superconductors can be understood on the basis of this 
hypothesis
\cite{pwa}, while others believe that any coupling between chains
must destroy the
Luttinger liquid.  One approach to this problem is
empirical: if evidence of Luttinger liquid behavior could be found for a 
real
weakly-coupled-chain system this would constitute
proof of the possibility of such behavior in higher dimensions.  In this
spirit, we ask whether $(TaSe_2)_4 I$ is, experimentally,
a Luttinger liquid, as has sometimes been claimed. 
As a first step, we compare ARPES data with a more conventional
interpretation, that of CDW fluctuations.
In a later paper, we shall attempt a more systematic 
comparison of strongly correlated electron and CDW theories
for the transport properties and the ARPES and core-level lineshapes
of $(TaSe_2)_4 I$.
  
In Secs.\ 2 through 4 of this paper 
we shall present the model used, its overall properties,
and the comparison of theory and experiment for ARPES lineshapes
in $(TaSe_4)_2I$.  The discussion is based on the 
premise that fluctuation effects born of the electron-phonon interaction 
dominate over those of correlation (interparticle interaction) 
for this system.  It is found that such a treatment works 
well for $(TaSe_2)_4 I$.  
A critical comparison of this picture with other candidate
descriptions is given in the conclusions
of Sec.\ 5.   

\section{The Model}

The problem of describing the combined effect of correlation and 
electron-phonon
interaction in one dimension on an equal footing is an axiomatically
hard one, since correlation in one dimension destroys the Fermi-Liquid 
picture on
which the (perturbative) treatment of electron-phonon interaction is 
based, whilst the
physics of the Peierls transition is dictated by $2k_f$ 
'backscattering' events,
which cannot be treated within the usual scheme for electronic 
correlations.  For a
given real material, in a given temperature range, however, it may not 
be necessary
to solve the general problem in order to understand the results of 
experiment.
As there is no clear framework for relating the microscopic properties
of materials to the parameters of {\it any} of the phenomenological 
models refered
to in this paper, the best that can be done is to develop a model based 
on what is
believed to be the relevant subset of physics for each material, and to 
test it
against a variety of experiments. 

In order to provide a framework for our calculation we briefly review 
here
some basic perturbative (mean field) results for electrons in a Peierls
CDW System, and present a succinct derivation of a Green's Function for 
electrons
suffering fluctuations of CDW order, in the spirit of  the treatment of 
Lee, Rice
and Anderson (hereafter referred to as LRA, \cite{lra}).

The starting point for any non-correlated theory of electron phonon
interaction is the Fr\"{o}hlich Hamiltonian
\begin{eqnarray}
H &= & \sum_{k} \epsilon (k) c^{\dagger}_k c_k
     + \sum_{q} \omega (q) b^{\dagger}_q b^{}_q
     + \frac{1}{\sqrt{L}} \sum_{q,k} g(q) c^{\dagger}_{k+q} c^{}_k u_q   
\label{eq:froh}
\end{eqnarray} 

\begin{eqnarray} 
u_q &= & \frac{1}{\sqrt{2\omega (q)}} ( b^{\dagger}_q +b_{-q} ),
\nonumber
\end{eqnarray}    
\noindent 
where $c^{\dagger}_{k}$ and $b^{\dagger}_{q}$ are (respectively) 
creation 
operators for 
electrons 
and phonons with dispersion $\epsilon (k)$ and $\omega (q)$, $u_q$ is 
the 
Fourier transform of 
the lattice displacement, and $g(q)$ the electron-ion coupling.  $L$
is the number of sites.  For simplicity  
we will not credit the electrons with spin, and will describe the band 
by a free 
particle 
dispersion relation, linearised about $k_f$.  

Consideration of the linear response of this system \cite{rice} reveals 
that the 
phonon frequencies 
undergo drastic adjustment if the band is near half filling; the $2k_f$
phonon is softened and its frequency vanishes at a well defined mean 
field 
transition 
temperature $T_c$ given by
\begin{equation}
\label{mftc} 
k_B T_c =  \frac{2\gamma}{\pi} e^{-\frac{1}{\lambda}} v_f k_f,  
\end{equation} 
with
\begin{equation} 
\lambda   =   N_0 \frac{|g(2k_F)|^2}{\hbar^2 \omega_{2k_F}^2},
\nonumber
\end{equation} 
\noindent
where $N_0$ is the density 
of states at the Fermi energy, and $\gamma \approx 0.5772$ is 
Euler's constant.  The overall picture of the 
important quantities in the mean field theory
is presented in 
Fig.\ 1, where the gap $\Delta(T)$ and the phonon frequency
$\omega_{2k_F}(T)$ are plotted on a phase diagram.

The 
result in Eq.\ \ref{mftc} (the giant Kohn effect) may also 
be found in low order perturbation theory, in which case a length 
scale 
$\xi_0$ emerges naturally from the calculation
\begin{eqnarray}
\label{mfxi} 
\xi_0 &= &  \frac{ \sqrt{7\zeta(3)} v_F }{4 \pi k_B T_c},  
\end{eqnarray}     
\noindent
$\zeta(3) \approx 1.202$ being the Riemann zeta function, and $v_F$ the 
Fermi 
velocity. 
Condensation of the 
softened $2k_F$ phonon leads to a static lattice distortion 
$<u_{2k_F}>$, and 
the opening of a 
gap $\Delta$ in the electronic spectrum.  The interaction term in the 
Fr\"ohlich 
Hamiltonian 
(\ref{eq:froh}) may then be approximated by a BCS-like mean field 
form  
\begin{eqnarray}
\label{bcshamiltonian} 
H_{ep}^{(MF)} &= & \sum_{k} [\Delta^* c^{\dagger}_{k-2k_F} c^{}_k 
                       + \Delta c^{\dagger}_{-k+2k_F} c^{}_{-k}]          
\end{eqnarray} 
                                                                                             
\begin{eqnarray} 
\Delta &= & \frac{1}{\sqrt{L}} g(2_{k_F}) <u_{2k_F}> .
\nonumber
\end{eqnarray}

\noindent
Self consistent solutions of $\Delta$ and $<u_{2k_F}>$ as a function of
temperature may now be 
found for the Peierls insulator in the same way as for a BCS 
superconductor.  
Both systems are 
dominated by the same square root singularity in the electronic density 
of 
states at the edges 
of the gap.

This picture of a metal-insulator transition in a purely one-dimensional 
metal 
is clearly not adequate 
as we expect fluctuations of the order parameter, when properly 
accounted for, to destroy the mean field 
solution at {\it any} 
finite temperature. Real systems, however, are not truly 
one-dimensional.  
$(TaSe_4)_2I$ 
comprises parallel chains of tantalum atoms \cite{gressier2}, 
surrounded by approximately perpendicular 
rectangles of $Se$ (Fig 2.).  The 
iodine resides between chains. Overlapping $d_z$ orbitals on the 
tantalum chains form a band along which conduction occurs.  Whilst 
electronic transport across the chains is believed to be diffusive at
the temperatures of interest here and thus does not give rise 
to coherent dispersion in the perpendicular directions, 
interchain interactions can act to stabilise the mean-field 
solution at some three-dimensional ordering temperature $T_c^{3D}$ 
considerably 
less than the mean 
field temperature $T_c$.  This then corresponds to the transition 
observed in real 
systems.  The LRA model of a one-dimensional metal does not explicitly 
include 
interchain effects, but 
their relevance and the three dimensional ordering temperature emerge 
very 
naturally from their analysis.  It is found that $T_c^{3D} \approx 
T_c/4$.  

The physical content of Lee, Rice and Anderson's extension of the 
mean field picture
is the realisation that the $2k_f$ fluctuations of the softened lattice 
are slow on
the timescale of electron dynamics and that they may therefore be 
approximated by a
static disorder potential, the determination of which is then a separate 
(classical) problem.

We start, then, from a natural generalisation of the BCS approximation 
to
the electron phonon interaction 
\begin{eqnarray}
\label{lrahamiltonian} 
\bar{H}_{ep} &= & \sum_{Q, k'>0} [ \Psi^*_{-Q} c^{\dagger}_{k'-Q} 
c^{}_{k'}  + \Psi_{Q} c^{\dagger}_{-k'+Q} 
c^{}_{-k'} ]  ,
\end{eqnarray}                                                           
where
\begin{eqnarray}
\label{orderparameter} 
\Psi_Q &= & \frac{1}{\sqrt{L}} g(Q) u_Q
\end{eqnarray}
are the components of the disorder potential and may be likened to an
order parameter for fluctuations of some portion of the lattice.  LRA 
then prescribe
an
equation of motion treatment of this hamiltonian, which after certain 
approximations
generates the following relation for the electronic Green's Function 
\begin{eqnarray}
\label{eqnforgreensfn}
{\cal{G}}(k,k;i\omega_n)^{-1} &= & \epsilon(k) - i\omega_n 
          - \sum_Q\Psi_{Q} \Psi^*_{-Q} \frac{1}{\epsilon (k-Q) - 
i\omega_n}   .
\end{eqnarray}
The same relation may be written down immediately by using the analogy with static
disorder simply to calculate the second order self energy correction for 
electrons scattered
by the potential $\Psi_{Q}$ \cite{Mahan}.  The Feynman diagram
used is shown in Fig.\ 3.  The box in the diagram is the phonon
self-energy, or the charge-charge correlation function
at $2k_f$.  The deformed 
lattice approximation treats the ionic configuration 
as rigid (incapable of recoil), with the positions given by a 
thermal average.  The absence of a  
frequency sum in
the elecron self energy expresses the fact that the lattice and CDW 
fluctuations have been
decoupled.

We note in passing that it would in principle be possible to substitute a 
Luttinger
Liquid Green's Function in the expression for the self energy, and so to 
treat the
effects of lattice distortion on a correlated system to a similar level 
of
approximation.  To obtain and parameterise spectral functions from such 
a
calculation would not, however, be trivial.  Similarly, refinements may 
be made to
allow for lattice recoil.  We shall not present these
here, but will limit our discussion to the Green's function previously 
found by LRA.

Evaluation of this expression then requires knowledge of the correlation 
function
for the lattice fluctuations at a given temperature.  A means of finding 
this was
supplied by Scalapino, Sears and Ferrell \cite{ssf}. They perform a 
functional
minimalisation of a generalised Ginzburg-Landau free energy to obtain 
correlation
functions for a fluctuating order parameter
in a one-dimensional system, and obtain the following form
\begin{eqnarray}
\label{correlationfn}
<\Psi(x)\Psi(x')> &= & <\Psi^2(T)>   \exp [-|x-x'|\xi^{-1}(T)]   
\cos[2k_F(x-x')]. 
\end{eqnarray}
LRA substitute a free energy with parameters taken from the mean field 
(linear
response)
perturbative treatment of the 1D Fr\"ohlich Hamiltonian : 
\begin{eqnarray}
\label{freeenergy}
F[\Psi_Q] &= & a(T)|\Psi_Q|^2 + b(T)|\Psi_Q|^4 + c(T)(Q - 
2k_F)^2|\Psi_Q|^2,  
\end{eqnarray}
with
\begin{eqnarray} 
a(T) &= & D_0 \frac{T - T_c}{T},
\nonumber
\end{eqnarray}
\begin{eqnarray} 
b(T) &= & D_0 [b_0 + (b_0 - b_1)\frac{T}{T_c}],
\nonumber
\end{eqnarray}
and
\begin{eqnarray} 
c(T) &= & D_0 \xi_0^2(T), 
\nonumber
\end{eqnarray}
\noindent
where $D_0$ is the (constant) density of states for the band, which is 
taken to 
have width 
$2\epsilon_F$. We fix $b_0$ and $b_1$ to give the correct zero 
temperature value 
of the gap 
$\Delta_0$ in the electronic spectrum: 
$b_0 =  \frac{1}{2\Delta_0^2}$
and
\begin{eqnarray} 
b_1 &= & b_0 \frac{7\zeta(3)}{16 \pi} \frac{(1.76)^2}{0.5}  . 
\nonumber
\end{eqnarray}

The problem of determining $<\Psi_Q^2(T)>$ and $\xi(T)$ then reduces to 
that of 
finding the low lying energy levels of a particle moving in an 
anharmonic 
potential well, the shape of which is determined by the coefficients of 
the free
energy 
\begin{eqnarray}
\label{anharmonicosc}
H &= & - \frac{1}{4} \frac{k_B^2 T_c^2}{D_0} 
\frac{\partial^2\Psi}{\partial x^2} 
           + a(T)|\Psi|^2 + b(T) |\Psi|^4
\end{eqnarray}

\noindent
This may solved numerically, or approximately using perturbation theory
and asymptotic analysis.   
It is found that the coherence length $\xi$ 
increases steadily from its mean-field value at $T_c$ 
with reducing temperature, but increases very rapidly at a 
temperature approximately one quarter of $T_c$.  
This 
implies that long-range order exists for temperatures 
below $\frac{1}{4}T_c$, and interchain coupling stabilises
the mean field 
solution.  We will identify this temperature
with the transition temperature $T_c^{3D}$ of 
a three-dimensional system 
and not attempt to treat interchain effects explicitly.  
The mean square value of the 
correlations increses approximately linearly 
with decreasing temperature, and takes on the role 
of a mean field gap below the three dimensional ordering temperature.

The following parameterisation is accurate in the temperature
range of experimental interest in the next section:
\begin{equation}
\xi^{-1}(T) =  \xi_0^{-1}(T) ( \frac{4T}{3T_c} - \frac{1}{3})
\label{xp1}
\end{equation}
\begin{equation}
<\Psi^2(T)> =  \frac{a'}{b} (1 - \frac{T}{T_c}) - \frac{1}{2} k_B =
\frac{T_c}{a'}                                            
\frac{1}{\sqrt{1-\frac{T}{T_c}}},
\label{xp2}
\end{equation}
where $a' =  a(T)/T $.

We are now in a position to assemble the Green's Function for the system, 
with the
(static) lattice fluctuations parameterised by the "gap" (squared) energy 
scale
$<\Psi^2(T)>$ and the "lifetime" energy scale $\xi^{-1}(T)$.
\begin{eqnarray}
\label{nearlygreensfn}
G_R (k,k;\omega)^{-1} &= & \omega - \epsilon (k)
          - \int dQ \frac{S(Q)<\Psi^2>}{\omega - \epsilon (k^+_-Q) + 
i\delta},
\end{eqnarray}
where S(Q) is a Lorentzian of width $\xi^{-1}$ centered on $Q=2k_F$, the sign in 
the denominator 
is chosen to give $ \epsilon (k^+_-Q) \sim \epsilon (k) $, and $<\Psi^2>$ is 
found from 
the results of SSF \cite{ssf}.  Evaluating the integral over Q, and dropping the 
second 
momentum index, we arrive at a result for the Green's function

\begin{eqnarray}
\label{greensfn}
G_R (k, \omega) &=& \frac{\omega + \epsilon (k) + iv_F\xi^{-1} }
                    {\omega^2 - \epsilon(k)^2 - <\psi^2> + iv_F\xi^{-1}(\omega - 
\epsilon(k)) }   .
\end{eqnarray} 
This will form the basis for most of the subsequent analysis.

\section{Basic Properties of the Model}

Insight into the properties of the model outlined may be obtained by 
consideration of the imaginary part of the Green's function 
(\ref{greensfn}) 
derived above: 
\begin{eqnarray}
A (k, \omega) &= & \frac{v_F \xi^{-1} <\psi^2>}
  {[\omega^2 - \epsilon(k)^2  - <\Psi^2>]^2 + v_F^2 \xi^{-2} [\omega - 
\epsilon(k)]^2}, 
\label{eq:spec}
\end{eqnarray} 

\noindent
where the parameters $<\Psi^2>$ and $\xi^{-1}$ have scale and 
temperature 
dependence given above.  As observed by LRA, this
may be integrated analytically to give an expression for the density of
states.  This is plotted for a system at temperature $T =  300K$=
with $T_c =  892K$, $T_c^{3D} =  263K$ and 
$\epsilon_F =  1.2 
eV$ in Fig.\ 4; the reason for this choice of parameters will be 
discussed in 
the light of photoemission data in a later section.  

The density of states shows clear evidence of a quasigap 
at all temperatures above $T_c^{3D}$ observable for a system such as 
$(TaSe_4)_2I$.  Spectral 
weight {\it is} still present at the Fermi energy ($\omega =  0$) at 
room 
temperature, but is greatly 
reduced.  Traces of the square root singularity which will dominate the 
mean 
field solution 
are visible at the edges of the gap for temperatures approaching 
$T_c^{3D}$.  
Sharp 
spikes and edges in the plot are a numerical artifact only. 

The LRA Green's function (\ref{greensfn}) 
above can clearly be seen to reduce to a BCS-type Green's 
function 
\begin{eqnarray}
\label{bcsG}
G_R (k, \omega) &= & \frac{\omega + \epsilon (k)}
                    {\omega^2 - \epsilon(k)^2 - <\psi^2> + i\delta },
\end{eqnarray} 

\noindent
with $<\Psi^2>$ taking on the role of a real gap in the limit where 
$\xi \rightarrow \infty$; this mean field solution is in fact the exact 
zero 
temperature limit 
of the LRA theory.  At the level of approximation relevant to the 
experiments the
model 
may be taken to posses a real gap
below $T_c^{3D}$.  We shall then proceed to describe the system by an 
LRA 
Green's 
function (\ref{greensfn}) above $T_c^{3D}$ ($263 K$ for $(TaSe_4)_2I$), 
and by a 
BCS Greens function below $T_c^{3D}$. 

It is also possible to integrate the spectral function numerically over 
$\omega$ 
to obtain a 
result for $n(k)$. (The integration my be performed analytically 
for the BCS-like expression below 
$T_c^{3D}$).  This is displayed in Fig.\ 5.  The occupation number is 
clearly 
dominated by the 
presence of the quasigap, varying over a scale in k-space given by 
$\Delta/\hbar v_F$.  No 
Fermi step is present in the occupation number at an experimentally 
observable 
temperature, but 
the region of k-space over which $n(k)$ undergoes most change becomes 
{\it 
smaller} with increasing 
temperature and decreasing size of the quasigap.  Both 
the occupation number and the density of states illustrate the fact 
that the metallic (fluctuating
CDW) phase of the model does not resemble a conventional Fermi liquid.

Consideration of the denominator of the spectral function at $k_F$ shows 
that
spectral 
weight is 
concentrated in one peak provided that the quantity 
$<\psi^2> - v_F^2\xi^{-2}/4$ is 
negative.  This gap like parameter will change sign at a temperature of 
similar 
magnitude to, 
but in general  different from, the mean field 
temperature $T_c$.  On cooling it becomes positive, 
and spectral weight is split into two peaks.  Since $T_c$ is 
well above room temperature, the
two-peak structure is the only one expected in the
experiments of the following section.
The breadth of the peaks is determined by the 
inverse coherence 
length $\xi^{-1}$.  At $k_F$ the division of spectral weight between 
peaks is 
even; for 
k-vectors deeper in the valence band more weight is found in the lower 
peak.  
This may be 
compared directly with coherence factor effects in the BCS-like phase of 
the 
model below 
$T_c^{3D}$.  The opening of the quasigap in the density of states is 
visible 
then in 
$A(k,\omega)$ as a splitting of spectral weight into two broad peaks, 
and the 
transition from 
metal to insulator marked by the progressive narrowing of these peaks 
until they 
become delta 
functions separated by a real gap at $T_c^{3D}$. Tight bunching of peaks 
near 
the edges of the 
gap lead to the pronounced rise in the density of states there; this 
will again 
resolve into a 
square-root divergence below $T_c^{3D}$.  

\section{Angle Resolved Photoemission}

Quasi-one-dimensional materials were chosen for  
angle-resolved photoemission experiments on 
account of the interesting phase transitions which 
they undergo, and also because the reduced
dimensional nature of their Brillouin zones simplified the 
interpretation of the spectra obtained.  The existence of many exactly 
soluble 
models of one-dimensional systems makes 
the exploration of quasi-one-dimensional materials 
equally appealing from a theoretical point of view.  
The particular hope of finding evidence 
for the existence of a Luttinger Liquid has motivated a large number of 
photoemission 
studies of quasi-one-dimensional conductors.  Among these,
the most detailed studies have been performed on $(TaSe_4)_2I$, 
\cite{dardel1,dardel2,hwu},
and the discussion of the data on this compound is our object in this 
section. 

While the authors of the papers on $(TaSe_2)_4I$ 
often disagree in the detailed interpretation of their data, 
the lack of spectral weight at the Fermi energy is a universally 
observed trend.  As this is one of the signal features 
of a Luttinger liquid, several researchers have concluded that 
$(TaSe_4)_2I$ and its sister compounds are Luttinger 
liquids in their conducting phase.  The 
nature of the loss of spectral weight may be best probed by angle 
resolved 
photoemission, since 
this offers some hope of establishing whither the missing weight has 
moved.

The quantity measured by an ideal angle-resolved photoemission 
experiment 
at zero temperature is the ground state
electronic spectral function of the material being probed.  This is 
formally
equivalent to the imaginary part of the system's retarded 
Green's function, and so confers complete knowledge of its single 
particle properties.  Real photoemission experiments suffer limitations
of finite temperature and resolution, and can only probe the properties
of the surface layer of material from which the electron emerges.  
Spectra 
also contain a substantial systematic background, which is not generally 
well 
understood, and which must be subtracted according to some prescription
before the data can be fully analysed. 

Recent angle-resolved studies of the conduction band in $(TaSe_2)_4I)$ 
have been performed above and 
below its CDW transition temperature, and 
the spectra possess a number of interesting features.  The one most 
stressed in the
literature is the lack of spectral 
weight at and near the Fermi energy.  Related to this is the 
lack of a Fermi step in the background of spectra taken{\it above} the 
transition 
temperature.  These features represent a marked departure from the usual 
character of metallic conduction, as observed in similar experiments on
three-dimensional systems.   
We shall see below that there are several other characteristics 
of the experimental results that are similarly anomalous.

The simplest result (non-interacting electrons) for energy-distribution
curves (EDC's) in ARPES done for initial electron momentum near the 
Fermi energy would be a delta-function.  Finite experimental resolution 
will
of course broaden this 
peak. 
We have convoluted our results with a Gaussian resolution
function.  The width is equal to the
published estimated resolution of 160 meV.  
One source of (not very interesting) background is that of
secondary electrons.  The intensity of this rises steeply
as the detected energy decreases, starting a few eV below
the Fermi energy.    
It was found that, for each spectrum, a Gaussian with width of similar 
order to 
that of the band and centred at an energy below the band
minimum could be chosen to closely mimic this contribution, and was duly 
subtracted.
The overlap of the Gaussian with the interesting structures
near the Fermi energy is small, but due to its width not quite 
negligible. 

As a rule, however,
the most striking difference between observed EDC's and the ideal
result is the existence of a structure resembling a Fermi edge
in addition to the expected peak.  This should be attributed to
the existence of quasi-elastic scattering of electrons on exit from the 
sample, 
probably arising from disorder near the surface.  It is important to 
remember that
electrons detected in ARPES originate from near the surface and few 
surfaces are
atomically flat.  To take this into account, it is most reasonable to 
suppose that
some fraction of the electrons have their momenta randomized on exit.

The resulting spectrum is a linear combination of a true
angle-resolved spectrum, (the spectral function),
and an angle-integrated spectrum (the density of states).
In theoretical terms, this means a combination of the 
imaginary part of the one-particle Greeen's function
and the imaginary part of its trace.  
This picture is confirmed by the observation that ARPES
EDC's and angle-integrated spectra taken on a single sample do 
differ only by the peak structure in the former.
The relative weight in the two components must be determined by a fit.

The experiments on $(TaSe_4)_2I$ fit this picture, with
one exception.  In contrast to
experiments on three-dimensional metals,
the angle-integrated spectrum near the Fermi energy
does not resemble the 
expected Fermi function.  Instead, the occupation falls 
off smoothly in the neighborhood of the chemical potential.
Qualitatively, this may be the result of Luttinger 
liquid behavior.  However, it may also result from the 
pseudogap in the LRA theory.  Only a quantitative comparison can
distinguish these alternatives.

We show the comparison of experiment and theory in Fig.\ 6.
Each plot is taken at a fixed angle of outcoming electron,
and the inferred wavevectors are as shown.
The data are taken from Ref.\ \cite{terrasi}.
The choice of wavevectors shown was dictated by the
availability of experimental data.
The background, fit by a Gaussian as mentioned above,
has already been subtracted from the experimental curves.
The theoretical curves are plotted from Eq.\ \ref{eq:spec}, 
broadened by convolution
with the Gaussian resolution function.  Two free parameters were 
retained
for the fit: the mean-field transition
temperature with a best-fit value $T_c =  892 K$ and the Fermi velocity 
$v_F =  6.5 \times 10^5 ms^{-1}$, which is fixed by the overall linear 
dispersion.  The parameter $k_F = 0.27 \AA^{-1}$ is fixed by the
fact that the band is one quarter filled.  This gives a zero temperature 
gap $\Delta_0 =  0.52 eV $ by extrapolation of the
dispersion to low temperatures.  $T_c$ compares well with the 
expectation that,
at a temperature around $T_c/4$, the actual transition
should take place and the resistivity should
become activated.  In $(TaSe_4)_2I$ this occurs at
$T_c^{3D} = 263 K$. 

Each plot shows a peak broader than the experimental
resolution of 160 meV.  The fits have been made with the
stated resolution.  It is evident that extremely good fits
could be made either by increasing the width of the 
convolving function by about 50 meV (30\%),
by assuming some scattering of
the electrons as they exit the material, or by assuming that the
material is impure to begin with, which would give
a momentum-independent additional width to the spectral function.  
Since all three of
these alternatives involve the introduction of an
{\it ad hoc} parameter, we have preferred to leave the theoretical
curves as shown and merely note that it would be somewhat
surprising if there was no source of broadening beyond
the instrumental resolution and that in Eq.\ \ref{eq:spec}.

Evaluating the comparisons in Fig.\ 6, we may say that the
peak positions are given very well.  The worst case is Fig.\ 6c, where 
the
theoretical prediction is too low by perhaps 30 meV, and the
other discrepancies are smaller.  The 
widths are too large by about 20\% in all cases, suggesting some
relatively minor additional systematic effect.

The momentum-integrated 
spectrum in Fig.\ 7 is obtained 
by integrating Eq.\ \ref{eq:spec} over $k$.  It may be compared                                                                 
with the results of a angle-integrated experiments \cite{dardel1}, 
\cite{dardel2}, and is 
clearly in good qualitative agreement with these, although the limited 
validity
of our linear approximation to the free electron dispersion renders 
quantative
comparision away from the Fermi energy impossible.
Most significant is the movement of weight away from the Fermi energy, 
signature of 
the pseudo gap caused by the charge fluctuations.   

\section{Fermi liquid, Luttinger liquid, or LRA liquid ?}

Four features of the LRA theory {\it and} the data are striking. \\
$(1)$  The movement of the peak position as a function of
momentum is very small near $k_F$.  In fact, the dispersion
relation, if it is defined as the peak position, apparently nears
a quadratic maximum at $k_F$. \\
$(2)$  The peaks are broad and symmetric.  The widths are not very 
momentum-dependent,
ranging only from about 400-600 meV in the range under study. \\
$(3)$  There is a strong pseudogap at all momenta, with the 
weight of the spectral function at the Fermi energy small.  In addition,
there is clear evidence of an energy scale associated with the
gap structure.  This is best read off from the peak position in Fig.\ 7
as about 500 meV.  This is related to the zero-temperature CDW gap in 
the 
LRA theory. \\
$(4)$  There are what may be called "shadow bands".  These are electronic
states, or rather peaks in the spectral function, where no bands should
be in a free-electron picture.  In the data, a clear peak is 
seen
even at momenta $|k| > k_F$.  These peaks shadow the ordinary band
in the range $|k| < k_F$ - they are translates of the ordinary
peaks through $\pm 2 k_F$.

On all four points the data agree qualitatively, and even
semiquantitatively, with the LRA theory.

Let us first compare these findings with the expectations of Fermi
liquid theory.  The spectral function should approach
\begin{equation}
A(k, \omega) \sim \delta(\omega - v_F (k-k_F))
\end{equation}
as $k$ approaches $k_F$.  Further from $k_F$, we expect
broadening due to interactions proportional to
$\omega^2$.  Specifically: \\
$(1)'$  In a Fermi liquid, the peak should disperse linearly
through the Fermi momentum. \\
$(2)'$  The peak should be symmetric, and the width
should be resolution-limited at $k_F$ and broaden away from
$k_F$. \\
$(3)'$  There is no gap or pseudogap and the only 
energy scales are $\epsilon_F > 1 eV$ and
$k_B T \approx 30 meV$. \\
$(4)'$  There are no peaks when $k > k_F$. \\
It is clear, comparing points $(1)'$ to $(4)'$ to
experiment $(1)$ to $(4)$, that this simple Fermi liquid behavior
is not at all consistent with the observations.

In the Luttinger liquid, we have a very different form 
for the spectral function at low energies.  In the spinless case
\cite{meden1}, the delta function characteristic of
the Fermi liquid is replaced by a power law singularity:
\begin{equation}
A(k, \omega) \sim \Theta( \omega + \tilde{v}_F|(k-k_F)|)
(- \omega + \tilde{v}_F(k-k_F))^{\gamma-1}
(- \omega - \tilde{v}_F(k-k_F))^{\gamma}.
\end{equation}
In this formula, $\Theta$ denotes the step function:
$\Theta(x) =  0$ for $x < 0$, and $\Theta(x) =  1$ for
$x>0$.  $\hbar \omega$ is the energy measured relative to the chemical
potential so that $\omega < 0$ for initial electron energies less than
the chemical potential.  $\gamma$ is the coupling 
strength for the electron-electron interaction.  For short (finite) 
range
interactions we expect $ 0 < \gamma < 1$, (the Infinite U Hubbard model
has $\gamma =  \frac{1}{8}$ \cite{1/8}).
Here $\tilde{v}_F$ is the excitation velocity, which includes 
contributions
from the kinetic energy and the interaction.  The integrated spectral
function (density of states) has the low-energy form:
\begin{equation}
\int A(k, \omega) dk \sim |\omega|^{2 \gamma}.
\end{equation}  
In the case of electrons with spin, there are generally
two singularities \cite{meden2}, one associated with the 
charge excitations at $\omega = \tilde{v}_{F,c} (k-k_F)$
and one associated with the spin excitations at
$\omega =  \tilde{v}_{F,s} (k-k_F)$.  If the velocities of the
two sorts of excitations are very similar, then at finite
resolution it may be difficult to distinguish this case from the 
spinless case.  We may now compare the Luttinger liquid scenario
to experiment. \\
$(1)''$  In a Luttinger liquid, for $\gamma < 1$ the singularity or 
singularities
disperse linearly through the Fermi energy.  For $\gamma > 1$, the 
dispersing
structure becomes very diffuse near the Fermi energy.\\
$(2)''$  The widths that should be observed in a 
real experiment are not universal - they depend on the 
details of the interaction \cite{meden3}.
One would expect some asymmetry in the peaks, however. \\
$(3)''$  There is a pseudogap-like feature in the 
density of states.  However, the exponent needed to
fit experiment is larger than expected \cite{meden3}.
There is no obvious energy scale in the theory
for repulsive interactions, though a peak may be observed in
the density of states for \cite{meden4} attractive interactions. \\
$(4)''$  There can be peaks for $\omega < 0$ and $k > k_F$.
However, they would be expected to be rather insignificant if
$0 < \gamma < 1$. \

Here it appears that $(1)''$, and probably the asymmetry
in $(2)''$ and the exponent in $(3)''$ are serious problems for the 
theory.

We conclude that the ARPES data on $(TaSe_4)_2I$ are not consistent
with either a Fermi liquid picture or a Luttinger liquid picture.

The most decisive observation is the failure of the peaks
in the EDC's to disperse through the Fermi energy.  The shapes and 
the widths of the peaks as a function of momentum are also difficult to
reconcile with these theories.  The picture of very strong
charge density fluctuations at $2 k_F$ appears, in contrast, 
to offer a consistent interpretation of all the data.We note, 
however, that 
the value of the CDW gap extracted from photoemission is larger than 
that obatined
from optical conductivity or resistivity measurements.

Efforts have been made in the last few years to obtain spectral 
predictions
for the Luther-Emery model \cite{voit}.  This provides a more 
natural framework
for consideration of the materials in question, since CDW fluctuations 
are not expected
to coexist with a Luttinger Liquid.  There is not yet, however, a 
universally agreed
prediction which may be compared with experiment, and the lack of a 
simple energy scale
in the model makes it difficult to extract quantitaive information from 
the real spectra.
It has also been brought to our attention since completion of this 
work, that a
more complete and self-consistent treatment of electron-phonon 
interaction has recently been
developed by McKenzie \cite{McK}.  A striking feature of his theory is 
the asymmetry of
the dispersing peaks in the spectral function, which is almost redolent 
of the Luttinger
Liquid, although quite different in its physical origin.  We have not 
attempted to make a
quantitative comparison of McKenzie's spectral function with the 
photoemission data for
$(TaSe_4)_2I$, but note that there is no observable asymmetry in the 
data.

It must be stressed that these conclusions apply to this
particular system only; experiments performed on other quasi-one-dimensional
systems suggest that the relative importance of
electron-electron and electron-phonon interaction effects is strongly
dependant on the system in question.  Even for $(TaSe_4)_2I$, 
correlation effects may be masked by the strong charge fluctuations only within a 
particular parameter regime; it is for example possible that the application of 
pressure to the chains would relatively strengthen the interaction effects.
In other materials, and particularly in some of the quasi 
one-dimensional SDW compounds, 
there would appear to be evidence for strongly one-dimensional 
correlation effects.

This work was supported by the National Science Foundation
under Grant No. DMR-9214739.  We would like to thank A. Chubukov, J. 
Voit, V. Meden, J. Ma, F. Mila, M. Onellion and 
C. Quitmann for useful discussions.  We are particularly grateful to 
the
authors of Ref.\ \cite{terrasi} for discussions of their data
and for supplying it to us in digitized form.

\Figures

\begin{figure}
\caption{Phases of a Peierls insulator  - from Kohn Metal with phonon
frequency $\Omega_2k_f$ to Band 
Insulator with gap $\Delta(T)$.}
\end{figure}

\begin{figure}
\caption{Schematic structure of the $Ta$ chain in $(TaSe_4)_2I$.}
\end{figure}

\begin{figure}
\caption{Feynman diagram for the electron self-energy
in the LRA theory, with the phonon self-energy indicated by the box.}
\end{figure}

\begin{figure}
\caption{LRA density of states as a function of energy and temperature
for the conducting phase.}
\end{figure}

\begin{figure}
\caption{Plot of n(k), the occupation number,
for temperatures above, below, and at (solid line) the 3D ordering 
temperature.} 
\end{figure}

\begin{figure}
\caption{Single parameter fits to ARPES spectra taken on $(TaSe_4)_2I$ 
in conducting phase, for various wavevectors for which data were available.} 
\end{figure}

\begin{figure}
\caption{Momentum integrated spectrum found from LRA model in conducting 
phase, showing suppression of the density of states at the chemical potential
$\mu = 0$.}
\end{figure}


\section*{References}

\begin{thebibliography}{40}

\bibitem{gressier1} P. Gressier, L. Guemas, A. Meerschaut,
                    Acta. Cryst. B {\bf 38}, 2877, (1982).

\bibitem{fujishita1} H. Fujishita, M. Sato and S. Hosino,
                     Solid State Comm. {\bf 49}, 313, (1984).

\bibitem{fujishita2} H. Fujishita, M. Sato, S. M. Shapiro and S. Hosino,
                      Physica B {\bf 143}, 201, (1986).

\bibitem{rocau} C. Rocau, R. Ayroles, P. Gressier and A. Meerschaut,
                 J. Phys. C {\bf 17}, 2293, (1984).

\bibitem{sato} E. Sato, K.Ohtake, R. Yammoto, M Doyama,
               T. Mori, K. Soda, S. Suga and K. Endo, 
               Solid State Comm. {\bf 55}, 1049, (1985).

\bibitem{geserich} H.P. Geserich, G. Scheiber, M. D\"{u}rrler, F. 
L\'{e}vy and 
P. Monceau, Physica B {\bf 143}, 198, (1986).

\bibitem{berner} D. Berner, G. Scheiber, A. Gaymann, H. M. Geserich, P. 
Monceau 
and F. Levy, Journal de Physique IV {\bf 3}, 255,(1993).

\bibitem{wang} Z. Z. Wang, M.C. Saint-Lager, P. Monceau, M. Renard,
               P. Gressier, A. Meerschaut, L. Guemas, J. Rouxel,
               Solid State Comm. {\bf 46}, 325, (1983).

\bibitem{johnston}  D. C. Johnston, M. Maki and G. Gr\"{u}ner,
                    Solid State Comm. {\bf 53}, 5, (1985).

\bibitem{dardel1} B. Dardel, D. Malterre, M. Grioni, P. Weibel and Y. 
Baer, Phys. Rev. Lett. {\bf 67}, 3144, (1991).

\bibitem{dardel2} B. Dardel, D. Malterre, M. Grioni, P. Weibel, Y. Baer, 
J. Voit 
and D. Jerome, Europhys. Lett. {\bf 24}, 687, (1993).

\bibitem{purdie} D. Purdie, I. R. Collins, H. Berger, G. Margaritondo 
and B. 
Reihl, Phys. Rev. B {\bf 50} 12222 (1994).
               
\bibitem{peierls} R. Peierls,
                  {\it Quantum Theory of Solids},
                  (Oxford, 1953).

\bibitem{rice} M. J. Rice and S. Str\"{a}ssler,
               Solid State. Comm {\bf 13}, 125, (1973).               
 
\bibitem{schultz} H. J. Schulz, 
                  Int. J. Mod. Phys. B {\bf 5}, 57, (1991).

\bibitem{pwa} P. W. Anderson and Y. Ren, in
              {\it Proceedings of the Los Alamos Symposium
              on High Temperature Superconductivity}, ed. 
              K. Bedell {\it et al.}, (Addison-Wesley, New York, 1990).

\bibitem{lra} P. Lee, T. M. Rice and P. W. Anderson,
              Phys. Rev. Lett. {\bf 31}, 463, (1973).

\bibitem{Mahan} G. D. Mahan, {\it Many Particle Physics}, p. 262 (Plenum, 
New York, 1990).
               
\bibitem{ssf} D. J. Scalapino, M. Sears and R. A. Ferrell,
              Phys. Rev. B {\bf 6}, 3409, (1972). 

\bibitem{gressier2} P. Gressier, M. H. Whangbo, A. Meerschaut and J. 
Rouxel, Inorg. Chem. {\bf 23}, 1221, (1984). 

\bibitem{terrasi} A. Terrasi, M. Marsi, H.Berger, G. Margaritondo, R. J. 
Kelly and M. Onellion, Phys. Rev. B {\bf 52}, 5592 (1995).

\bibitem{hwu} Y. Hwu, P. Almeras, M. Marsi, H. Berger, F. Levy, M. 
Grioni, D. Malterre and G, Margaritondo,  Phys. Rev. B {\bf 46}, 13 624, (1992).

\bibitem{meden1} A. Theuman, J. Math. Phys. {\bf 8}, 2460 (1967).;
A. Theuman, Phys. Lett. A {\bf 59}, 99 (1976).; A. Luther and I. Peschel, 
Phys. Rev B {\bf 9}, 2911 (1974).; F.D.M. Haldane, J. Phys. C {\bf 14}, 2585 (1981).

\bibitem{1/8} M. Ogata and H. Shiba, Phys. Rev. B {\bf 41}, 2326 (1990).;
              A. Parola and S. Sorrela, Phys. Rev. Lett. {\bf 64}, 1831 (1990).;
              H. Schultz, Phys. Rev. Lett. {\bf 64}, 2831 (1990).

\bibitem{meden2} V. Meden and K. Sch\"onhammer, Phys. Rev. B {\bf 46}, 
1573 (1992).; J. Voit, Phys. Rev B {\bf 47}, 6740 (1993).; J. Voit, J. 
Phys. C {\bf 5}, 8305 (1993).

\bibitem{meden3} K. Sch\"onhammer and V. Meden, Phys. Rev B {\bf 47}, 
16205 (1993).

\bibitem{meden4} K. Sch\"onhammer and V. Meden, J. Elect. Spectr. Relat. 
Phenom. {\bf 62}, 225 (1993). 

\bibitem{voit} J. Voit, V. Meden, (private communications).

\bibitem{McK} R. H. McKenzie, Phys. Rev. B, {\bf 52}, 16428 (1995).

\end{thebibliography}
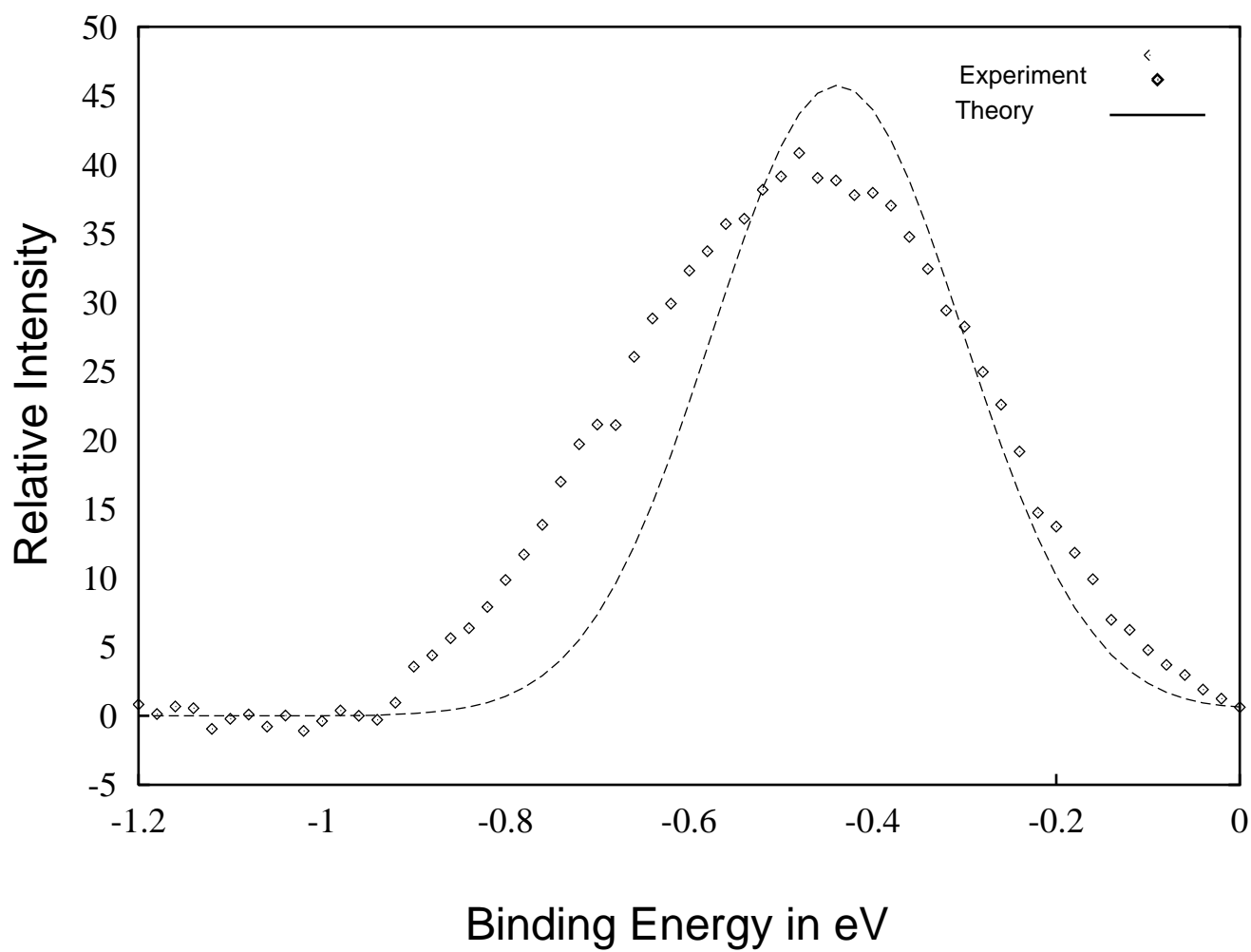
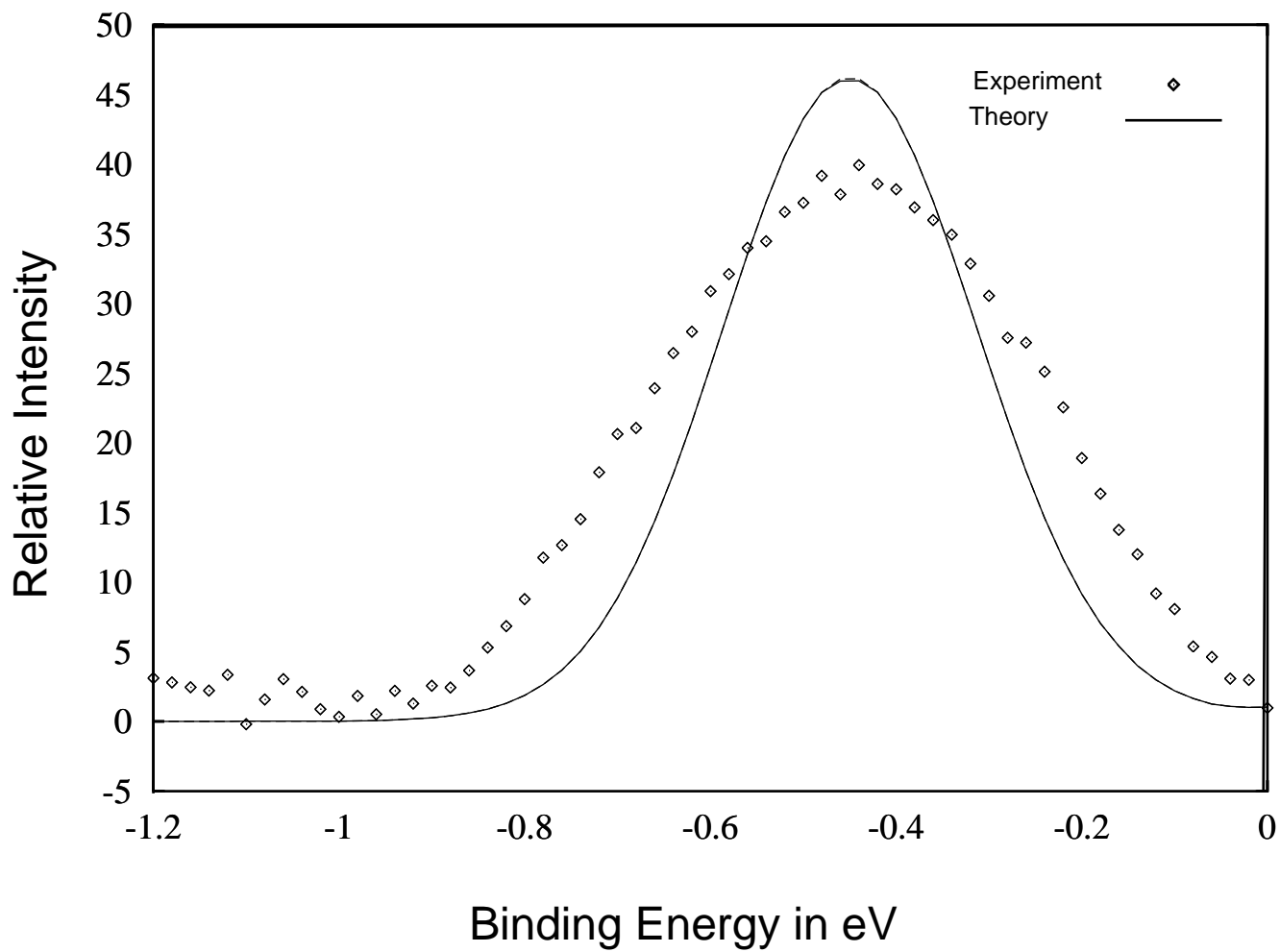
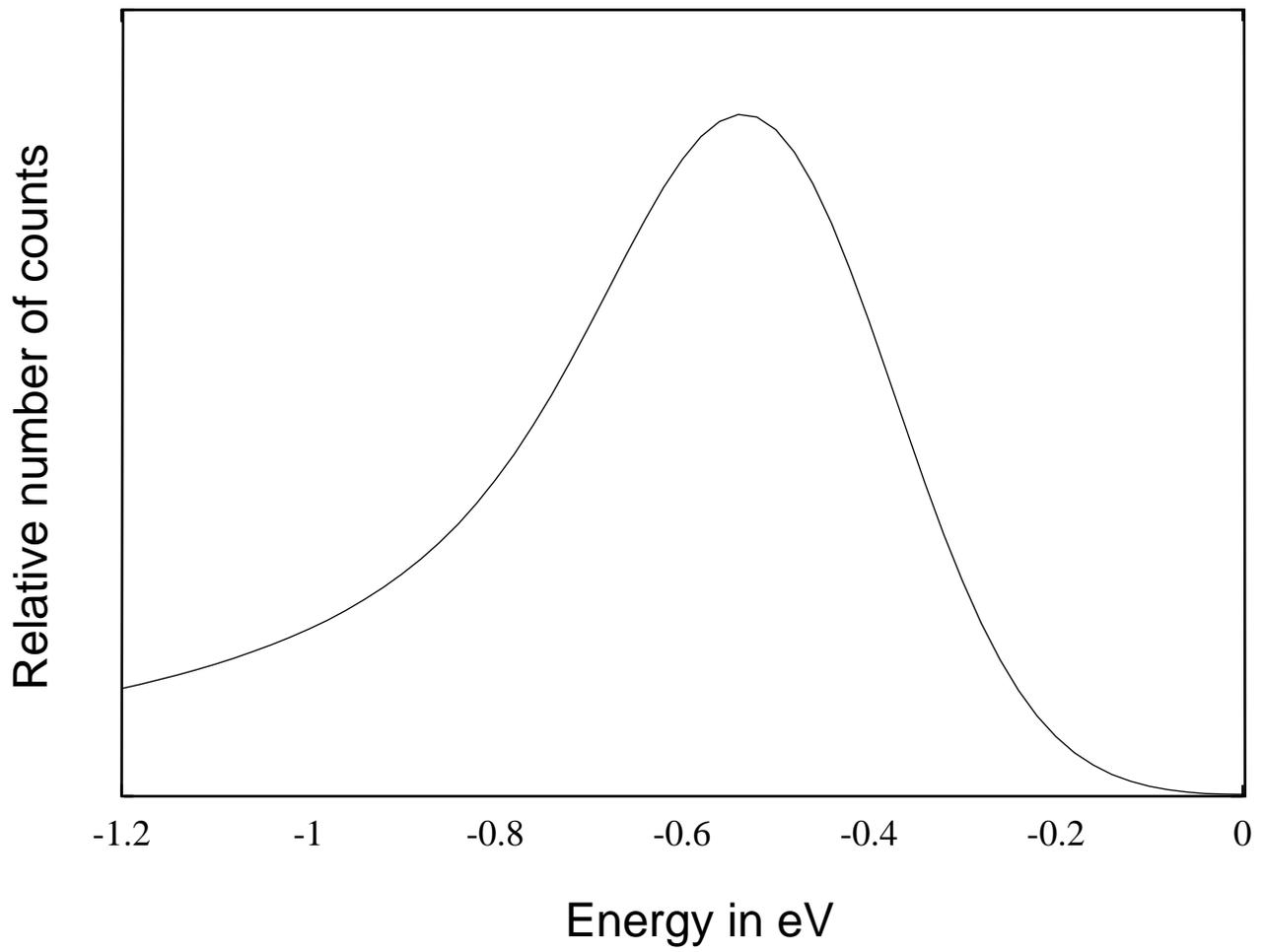
\end{document}